\documentclass[letterpaper,twocolumn,10pt]{article}
\usepackage{usenix2021_SOUPS}
\microtypecontext{spacing=nonfrench}

\usepackage{tabularx}
\usepackage{graphicx}

\usepackage{tikz}
\usepackage{amsmath}
\usepackage{color, colortbl}
\usepackage{csquotes}

\definecolor{Gray}{gray}{0.9}

\newcommand\removedForSpace[1]{{{{}}}}

\newcommand{\eg}{e.\,g.}

\newcommand{\contextquote}[2][]{{{\emph{``{#2}''}\,---\,[#1]}}}

\begin{document}

\date{}

\title{\Large \bf \enquote{What Keeps People Secure is That They Met The Security Team}:\\ Deconstructing Drivers And Goals of Organizational Security Awareness}

\def\plainauthor{Hielscher \& Parkin}

\author{
{\rm Jonas Hielscher}\\
Human-Centred Security \\
Ruhr University Bochum, Germany
\and
{\rm Simon Parkin}\\
Cybersecurity (Technology, Policy, and Management)\\
Delft University of Technology, Netherlands
}

\maketitle\thispagestyle{empty}
\pagenumbering{gobble}

\begin{abstract}
Security awareness campaigns in organizations now collectively cost billions of dollars annually. There is increasing focus on ensuring certain security behaviors among employees. On the surface, this would imply a user-centered view of security in organizations. Despite this, the basis of what security awareness managers do and what decides this are unclear. We conducted $n=15$ semi-structured interviews with full-time security awareness managers, with experience across various national and international companies in European countries, with thousands of employees. Through thematic analysis, we identify that success in awareness management is fragile while having the potential to improve; there are a range of restrictions, and mismatched drivers and goals for security awareness, affecting how it is structured, delivered, measured, and improved. We find that security awareness as a practice is underspecified, and split between messaging around secure behaviors and connecting to employees, with a lack of recognition for the measures that awareness managers regard as important. We discuss ways forward, including alternative indicators of success, and security usability advocacy for employees.
\end{abstract}

\section{Introduction}
Research has made great strides in informing security awareness practice and awareness campaigns~\cite{jayatilaka2021evaluation,Khando.2021,bada2019cyber,aldawood2019reviewing}, in parallel with increased interest in the human factors of security over the past two decades and more. This has resulted in a greater understanding of awareness campaigns themselves, but not necessarily the drivers for how they are managed. 

Recent research has evidenced that those managing employee security in organizations may have motives and measures outside of secure working itself~\cite{Hielscher.2023a,bada2019cyber}, such as regulatory compliance. A disconnect may also exist between managers and the users they are looking after ~\cite{ashenden2016security,ashenden2013cisos,reinfelder2019security,dasilva2022cyber,Hielscher.2023a,Hielscher.2023b}. In 1999, Adams \& Sasse characterized employees not using provisioned security apparatus in the way the organization had anticipated~\cite{adams1999users}. This raises questions as to the incentive structure and mechanisms of change, that would drive adoption of improved practices as derived in prior research. It is then natural that in recent years, guiding employees to behave securely has grown to be a business function in its own right -- worth billions~\cite{VMR.2023}.
A heavy focus of work in this space has been on designing effective interventions focusing on secure behavior, but less on how these interventions are effectively positioned within the specific setting of an organization.

Prior work has illustrated how drivers and goals, especially relating to security, may differ between employees~\cite{beautement2008compliance} and the managers of the system. Research up to now has focused on what the apparatus allows an awareness campaign to achieve -- where it has been questioned whether security awareness campaigns in organizations exist only to achieve security awareness~\cite{bada2019cyber}. Less attention has been given to the decisions and drivers that shape what is possible for a Security Awareness Manager (SAM) to achieve in their role~\cite{demjaha2022boundedly}, though there has been work exploring the requirements for \eg, US local government awareness campaigns to succeed~\cite{haney2022approaches}. This prompts a need to determine if secure working behavior is the only driver influencing the activities of a SAM, and in turn, security awareness programs.

SAMs are those members of an organization tasked with managing the security awareness of employees, typically by conducting awareness campaigns or buying and managing awareness products (like phishing simulations) from external vendors. In most cases, regular members of the (technical) security teams, or Information Security Officers, are working as SAMs part-time~\cite{SANS.2021.AwarenessManagerReport,SANS.2022.AwarenessManagerReport}. In larger organizations full-time SAMs are becoming more common -- like all participants in our study.

Here we study how security awareness activities are perceived by practitioners whose job it is to manage security awareness in (large) organizations -- the SAMs. We explore their aims, how they demonstrate success, and -- ultimately -- how they sum their role with what is expected of employees in the same organizations. We identify the drivers and influences over their decisions, and to what extent this aligns with the goals of behavior change and behavior support (such as smoothing the path to effective change \cite{fogg2019tiny,clear2018atomic}).
To do so, we performed semi-structured interviews with $n=15$ European SAMs, all working primarily in awareness and for large public and private organizations. SAMs are the primary gatekeepers for any change and improvement of real-world security awareness. Beyond a specific examination of US local government efforts to ensure employees are trained (see~\cite{haney2022approaches,jacobs2021exploring}), to the best of our knowledge, no previous in-depth research has studied this group's perception. 

We formulate the following \hypertarget{researchQuestions}{research questions}: 

\begin{itemize}
    \item \textbf{RQ1}: What activities and topics do security awareness managers regard as being security awareness, within the remit of their role? 
    \item \textbf{RQ2}: How do security awareness managers interact with employees?
    \item \textbf{RQ3}: How is success defined for security awareness managers, by them or others?
\end{itemize}

\vspace{0.3cm}

Looking at security awareness through the lens of SAMs and their perspective, we identify a range of critical outcomes. These include that, according to our participant SAMs, security awareness has a goal of ensuring active engagement with security; this is then not a direct measurement of increased security levels through behavior change. The positive feedback from employees is then used as a success indicator. Also, security awareness is both an underdefined and overstretched term, used to refer variously to training, communication, human factor principles, and usable security. Our suggestions to move forward include a clearer, modified mandate for SAMs (as employees' security advocates), the development of clear definitions around security awareness and related concepts, and research that takes into account the SAMs' struggle with raising engagement itself.

\paragraph{Contribution:} 
(I) We study security awareness through the lens of practitioners from large European organizations, characterizing dependencies and drivers for their decisions around how security awareness appears for employees;
(II) We identify barriers to improving awareness programs. 
We highlight how awareness is currently locked to a specific path, where usability is defined by vendors, and SAMs have limited tools for responding to engaged employees who cannot follow the training that the SAMs offer;
(III) We address the under-specification of awareness activities in research, beyond an expectation of structured, regular training (where there are also ad-hoc advisories driven by technical teams, for instance).

\section{Background \& Related Work}
Here we explore the definition of security awareness in literature (Section~\ref{sec:back:whatIsAwareness}), what its goals are (Section~\ref{sec:back:aims}), challenges related to how awareness happens in practice in organizations (Section~\ref{sec:back:awareInOrgs}), and the SAM role (Section~\ref{sec:back:SAM}).

\subsection{What is Security Awareness}
\label{sec:back:whatIsAwareness}
As a foundation, we refer to US NIST standards -- NIST standard 800-12 defined employee-facing \enquote{awareness, training, and education} (ATE) for security, in the mid-nineties~\cite{nist800-12-1996}; the what, how, and why of security behavior respectively. The NIST 800-50 standard appeared later in 2003, focusing on building a dedicated program for security awareness~\cite{nist800-50-2003}. They define awareness as a process to focus attention, training to generate necessary skills, and education to integrate all skills. Here, learning is described as a continuum that starts with awareness.
In reality, the three components of awareness, education, and training are often referred to collectively by the shorthand of \enquote{security awareness}, and are considered to be hard to distinguish in security research~\cite{Amankwa.2014}.

Prominent issues covered by security awareness activities include the creation of secure passwords~\cite{eminaugaouglu2009positive}, and advice for how to detect phishing emails~\cite{lain2022phishing,Brunken.2023,carella2017impact}. There have also been trends in the field, some of which are ongoing, such as \eg, \enquote{security culture}~\cite{da2010framework,renaud2014curious}.
Looking ahead to our interviews, we observe that security awareness is more recently being referred to in practice as security behavior change -- another trend in the field, also promoted by the US-based SANS institute.\footnote{Security Awareness Maturity Model - Promoting Change: \url{https://www.sans.org/blog/security-awareness-maturity-model-promoting-change/}, accessed \today}

Regarding the positioning of security awareness activities within organizations, there will generally be computing devices and services provided by the IT team, complemented with a policy or declaration of expected ways to behave and use those provisions~\cite{Kirlappos.2014}. Employees are then expected to follow the rules and use the provided tools -- to not do so has in the past been regarded as willing non-compliance~\cite{adams1999users}, but in recent years there has been a growing body of research into users' knowledge and capacity to engage (or not) with security, \eg, with policy non-compliance \cite{beautement2016productive}, and employee motivations to use what is provided \cite{blythe2015unpacking}.

Informational and training materials may be deployed as \eg, online training and security alerts, but also instructor-led events and engaging artifacts such as posters and branded items~\cite{abawajy2014user,bauer2013end} (as has also been advocated in prior NIST standards such as 800-50).

\subsection{Goals of Security Awareness}
\label{sec:back:aims}
Looking broadly at what secure behavior is, there have been attempts to standardize what good behavior is and how to measure it, as with SeBIS~\cite{egelman2016behavior}, which focuses on general behaviors such as having a secure password and security checks when online. The HAIS-Q~\cite{parsons2017human} instrument has a focus on organizations, including not only secure practices but awareness and alignment with IT provisions (including policies and compliance with them). The focus of these works is generally to determine how well individuals are engaging with and applying (provisioned) security solutions. 

Looking at industry norms we see that, for example, ISO 27004~\cite{iso27004.2016} explains that organizations should measure whether employees are prepared against social engineering through phishing simulation click-rates. The BSI basic protection~\cite{bsi.2022} states that the goal of awareness campaigns would be to raise employees' awareness of security risks. In practice, awareness campaigns often require the recipient to already have a lot of existing security-related skills -- in order for expected behaviors to take root~\cite{bada2019cyber} -- and to also commit sizeable effort to engage. Further, progress is generally not adequately measured (in terms of relating awareness activities to what employees actually do in practice). Here we explore how restrictions in guidance influence SAM activities.

\subsection{Awareness in Organizations}
\label{sec:back:awareInOrgs}

The influence of factors outside of security behavior has been explored elsewhere, for instance through the lens of policy creation \cite{cram2017organizational}, and how organizational factors influence what goes into those policies. Here we explore incentives and constraints imposed upon the SAM who must support the security behavior expectations that those factors inform. Pallas \cite{pallas2009information} applied an economics perspective to understand how security is managed in organizations, considering the pros and cons of using hard and soft controls and other related enforcement costs and resource needs (finding, for instance, that soft human-facing controls such as policies require comparatively costly, follow-up enforcement). 

Focusing on employee-facing controls, specifically anti-phishing solutions, Brunken et al.~\cite{Brunken.2023} found that multiple stakeholders were involved, and that alignment between them all was difficult to find. Such a security solution could, for instance, create a clash between the security team and the helpdesk, if an increase in support queries is anticipated, triggered by anti-phishing measures.
Taking a wider organization-level view, Moore et al.~\cite{moore2016identifying} examined the incentives and drivers for CISOs, finding that upper-management support and regulatory alignment were key influences, and with this, that the freedom to define the role and activities varied across security professionals and organizations. 

The question, as to what exactly security awareness is in practice, has been asked before~\cite{scholl2018scientific,Hansch.2014}. We examine this further now, to understand how aspects of security awareness relate to security management efforts in organizations.

\subsection{Security Awareness Managers}
\label{sec:back:SAM}
Security Awareness Managers (SAM) are tasked with raising security awareness among employees, regardless of their place in the organization. Typically this is done through the development or implementation of campaigns and training of different forms. However, no uniform definition for SAMs exists yet, all while \eg, the SANS institute or the German TUV offer training courses to become a certified SAM. Here the TUV envisions a SAM as someone who coordinates and strategically plans security awareness activities in their organizations.\footnote{Security-Awareness-Koordinator [German]: \url{https://akademie.tuv.com/weiterbildungen/security-awareness-koordinator-tuev-473465}, accessed \today}

Typically, a SAM is an employee in the field of security who manages security awareness as one duty alongside another dedicated role~\cite{SANS.2021.AwarenessManagerReport,SANS.2022.AwarenessManagerReport}; it may be that over 80\% of security awareness professionals spend half or less of their time on awareness, given this model. 

It was found that, among awareness managers in U.S. government agencies, approximately 90\% manage awareness in a part-time capacity, with 55\% managing awareness in less than 25\% of their time~\cite{Haney.2022}. In some cases, even a CISO, or an ISO27001 Information Security Officer~\cite{iso27004.2016} might act as a SAM. Especially in larger organizations, there might be multiple SAMs, working in an awareness team.
There are few other studies directly focusing on the experiences of SAMs outside of the materials they produce: Haney et al.~\cite{haney2022approaches} studied U.S. governmental SAMs through focus groups and a survey. Among other things, they were interested in SAMs' collaborations, the importance of regulations for their work, and potential measurements. In an associated study, Jacobs et al.~\cite{jacobs2021exploring} found that effectiveness was measured through training engagement and event attendance, noting challenges in resourcing the materials needed for employees to know about secure practices. Blythe et al.~\cite{Blythe.2020} surveyed 98 security awareness professionals to understand their perception of sanctions for non-compliant behavior.

It is important to note that security and privacy awareness in organizations is not necessarily always managed centrally. So-called security and privacy champions/ambassadors/advocates -- employees willing to voluntarily promote secure behavior in their teams, often without a direct formal mandate from the security managers -- are spreading their knowledge to raise awareness among their colleagues~\cite{tahaei2021privacy,haney2017skills,haney2017work,haney2018itsdull,Menges.2023.CaringNotScaring,gabriel2011selecting,becker2017finding}.
Looking ahead to our findings, those champions were welcome allies for the work of our participants (see Section~\ref{sec:res:employees}).

\section{Methodology}
\begin{figure}[]
	\centering
	\includegraphics[width=8.8cm]{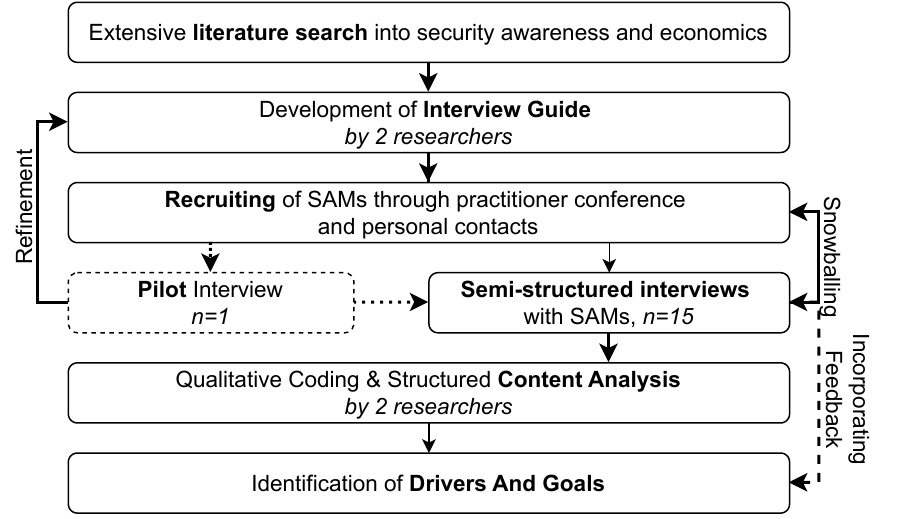}
    \caption{Our data collection and analysis methodology.}
	\label{fig:method}
\end{figure}

We conducted semi-structured online interviews with $n=15$ security awareness managers (SAMs). The interviews were carried out from May to September 2023. Figure~\ref{fig:method} summarizes our method.

\subsection{Instrument Development}
The development of our interview guide was informed by the literature review we performed on the practice of security awareness in organizations. The interview guide covered: (I) the SAM role; (II) activities related to this role; (III) goals of awareness for the SAMs and their organizations; (IV) awareness tools \& techniques; (V) the role of employees in security awareness, and; (VI) an outro and debrief. 

Regarding the management of awareness as a role, this informed questions regarding the availability (or scarcity) of resources to support the role (which can lead to trade-offs \cite{parkin2010stealth}), and decision-making drivers and influences (including whether the remit of the role or colleagues in the organization encourage particular decisions). We then have \textit{role} questions and \textit{role activities} questions.

Our position is that the majority of research focuses on behavior changes in isolation, and not how they are situated as a function of business, alongside other interests. Prior research has identified clashes between primary and secondary tasks in security~\cite{flechais2005divide,herley2009so}, and in organizations (\eg,~\cite{beautement2016productive}). In essence, managing awareness is a job, with pressures, and an activity that must sit alongside other non-security activities~\cite{herley2013more}, that uses infrastructure provisioned by the organization.

For these reasons, we wanted to understand the \textit{goals of awareness}, what \textit{awareness tools \& techniques} are available, and their suitability. It is only after establishing the empowerment of the role of a SAM, that we position this alongside the \textit{role of employees in security awareness}. Prior research with CISOs~\cite{Hielscher.2023a} has identified a range of CISO perspectives, for instance, ranging from a view that employees simply have to find a way to do what is expected for security (because it is important), to the CISO struggling to find a way to support employees to work but be able to do so securely.

Our initial scope was to identify incentives, and perverse incentives~\cite{anderson2001information} as well as misaligned incentives~\cite{anderson2006economics}; these mechanisms can help to explain observed (mis)alignment of drivers. Where a combination of economics with psychology generally points to behavioral economics~\cite{anderson2009information}, we took a step back to look at the infrastructure around behaviors, that defines and limits which behaviors are deemed acceptable -- and not -- within a system (in this case, an organization). From here we derived our research questions. 

A pilot interview with a SAM of a German industrial organization led to further refinements of the interview guide, mainly in the form of a reduction of the question set to stay within time constraints. The results of this pilot interview were not included in the final analysis. The full interview guide can be found in Appendix~\ref{app:InterviewGuide}. 

\subsection{Recruitment}
We only recruited SAMs that manage awareness as their primary job function. We hypothesized that organizations that have such dedicated SAMs implement \textit{industry awareness best practices}. Hence, studying their SAMs can give insights into the state-of-the-art awareness implementations in practice that other, smaller organizations may assume to mimic for best results.
Recruiting such a number of primary job SAMs was challenging and it took us from May till September 2023 to complete the recruiting. SAMs that are primarily employed for awareness tasks are a difficult-to-reach population (as with other security professionals such as CISOs~\cite{reinfelder2019security}).

An initial recruitment channel was a security awareness practitioner conference held in Germany. Here one of the authors
approached various participants of the conference and directly invited them to the interviews. This was combined with the recruitment of two SAMs via personal contacts, and snowballing to reach the remaining participants.

Three SAMs, who initially agreed to interviews, dropped out due to time constraints, resulting in 15 interviews in total (excluding the one pilot interview).

We did not compensate the participants financially, but instead offered each participant early sight of the study results before publication. When we sent out the results we also actively asked for participants' feedback on the content. Nine interviews were conducted in German, and six in English.

\subsection{Analysis}
The interviews were automatically transcribed with Open AI's whisper program\footnote{Whisper: \url{https://openai.com/research/whisper}, accessed \today} (offline, on the local machine, so no data was sent OpenAI) and then manually corrected by one researcher.
We used MaxQDA 2022 for the coding process. We applied Kuckartz’~\cite{kuckartz.2012} process scheme of content-structuring analysis and followed Braun \& Clarke's~\cite{clarke2015thematic,braun2021one} method of thematic analysis, where we implemented \enquote{codebook style} coding. Here we applied combining deductive and inductive coding strategies and a category-based evaluation along main codes.

Two researchers were involved in the analysis, with only one researcher directly coding in the interviews, due to only this researcher being able to understand German. We deemed this the better alternative, to coding translated documents, where meaning might have gotten lost. Following Braun \& Clarke's~\cite{clarke2015thematic,braun2021one} guidelines, one coder is sufficient in the case of \enquote{codebook style} coding~\cite{McDonald.2019}. The second researcher shaped the coding in all steps through regular meetings (at a regularity of days rather than weeks) and direct discussion of the codes, resulting in adding/removing/updating of codes, toward the identified themes. 

Additionally, both researchers created memos (summaries) after every interview session as an interim step before the systematic coding, which started after the first nine interviews were completed, where they would directly discuss the impression that the interviewer had on the interviews. Hence, the coding process had to be guided by multiple discussions (dozens of hours of joint sessions), where the main coder presented their coding strategy based on translated examples to coder two. Those discussions led to refinements in all steps of the coding process. 

The coding happened in multiple steps: (I) A deductive codebook was created based on the interview guide and the research questions. (II) All interviews were coded deductively and inductively. (III) The code book was refined. (IV) All interviews were coded with the finalized codebook. During all steps, memos were created that guided the discussion. 
The codebook can be found in our replication package\footnote{\url{https://doi.org/10.4121/9dc01aa6-8274-43f4-b137-6d185e7008d1  }}, with a mapping to where themes are discussed in the Results section.

\subsection{Ethics and Data Privacy}
Our study was reviewed and approved by the ethics committee at TU Delft. Ruhr University does not have an institutional review board (IRB) nor an ethics review board (ERB) for security research. We followed the Menlo Report principles~\cite{kenneally2012menlo} and GDPR requirements: all participants received a data privacy statement prior to the interviews. We informed the participants about their rights, especially that the interviews were voluntary and could be stopped at any time, and that we would only publish anonymous results of the interviews. All participants gave their explicit consent. We recorded the audio data of the interviews and ensured to delete this data after six months. The data was exclusively stored at our institutions, within the jurisdiction of the GDPR. Due to the small size of the SAMs community, we only report demographic data in an accumulated form to prevent identification.

\subsection{Limitations}
As with every study with human subjects, our research has limitations. Firstly, the participants might have held self-reporting and social desirability biases and might have kept information from us. This may reflect, for instance, how although our participants noted frustrations, they did not comment on any failure in their roles and activities.

The codebook reached saturation during the first 9 interviews (with all participants from central Europe) at which point we felt it conducive to explore other regions. Interviews with the UK SAMs brought new angles, \eg, a usable security perspective (see Section~\ref{results:securityfriction}). In summary, we did not reach saturation. With a much larger set of participants from the same regions this would have been possible, but due to the nature of this hard-to-recruit population, which would have added considerably to the approximately six months required for the recruitment of the existing 15 participants, we refrained from further recruitment.
Here we look only at large organizations employing someone in a `security awareness' role. Smaller organizations face many challenges around security awareness centered around a lack of resources~\cite{bada2019developing}.
Our SAMs came exclusively from European countries.

\section{Results}
Here we first summarize the demographics of our participants, before we present the results of our content analysis.

\subsection{Demographics}
All participants worked in security awareness as their primary job function. Twelve participants were actively working as SAMs, and three SAMs transitioned from working inside of organizations to awareness consultant roles recently. The SAMs had diverse backgrounds like engineering, computer science, psychology, education, and marketing. Participants 1-9 worked in the \textit{DACH-region} (Germany, Austria, Switzerland), P10 in Denmark (previously in the UK), and P11-15 in the UK. At least three SAMs have worked in different European countries in an awareness role. 
Table~\ref{tab:demographics} summarizes key properties of the participants. 

While our analysis did not aim to understand differences between regions, or industries, we identified a few systematic differences, namely:
(I) the different interpretations of employees' roles between the DACH managers and all others (See Section~\ref{sec:res:employees}),
(II) the focus on sector-specific regulations (see Section~\ref{results:regulations}).

\begin{table}[]
\small
\caption{Background information of the SAMs.}
\label{tab:demographics}
\begin{tabularx}{8.5cm}{lXX|lXX}
\hline
\hline
\textbf{Gender}                                        & \#               & \%          & \textbf{Sector}           & \#                   & \%                   \\ \hline
Female                                                 & 10               & \textit{66} & Energy           & 3                    & \textit{20}          \\
Male                                                   & 5                & \textit{33} & Consulting       & 3                    & \textit{20}          \\ \cline{1-3}
\multicolumn{3}{l|}{\textbf{Education}}                                                           & Banking          & 3                    & \textit{20}          \\ \cline{1-3}
Cyber \& Inf. Sec                                      & 5                & \textit{33} & Industry         & 2                    & \textit{13}          \\
CS \& Engineering                                      & 4                & \textit{26} & Retail           & 2                    & \textit{13}          \\
Comm. \& Marketing                                     & 2                & \textit{13} & Public           & 1                    & \textit{6}           \\
Social Science                                         & 2                & \textit{13}  & Automotive       & 1                    & \textit{6}           \\ \cline{4-6} 
   Education                                              & 1             & \textit{6} & \textbf{Country} & \multicolumn{1}{l}{} & \multicolumn{1}{l}{} \\ \cline{4-6} 
Psychology                                             & 1                & \textit{6}  & UK               & 5                    & \textit{33}          \\ \cline{1-3}
\multicolumn{3}{l|}{\textbf{Number of Employees}}                                        & Germany          & 4                    & \textit{26}          \\ \cline{1-3}
\textit{Max}:                400,000   \& \textit{Min}:  1,600   & & & Switzerland      & 4                    & \textit{26}          \\
 \textit{Median}: 27,000 &  &  & Austria          & 1                    & \textit{6}           \\
\textit{Average}: 62,000                                                      &                  &             & Denmark          & 1                    & \textit{6}    \\
\hline
\hline
\end{tabularx}
\end{table}

u\subsection{Awareness in Organizational Practice}
The concepts of awareness are shaped by the organizations they are carried out in~\cite{jacobs2021exploring}, which we explore in this section.

\subsubsection{Activities}
\label{results:activities}
Awareness is implemented as a multitude of activities for an organization's employees. For this, the SAMs combine e-learning, in-person events, simulations, and direct communication. Scalability of the activity was a prime argument for digital solutions, accepting that they might be less successful.

Phishing simulations were at the core of most SAMs' (P1,2,4-P10,12,15) awareness activities, as seen elsewhere~\cite{haney2022approaches}. Here we got an insight into why this might be the case: they are just best practices the SAMs see in other organizations (P8,15), regulations require them (P6,9,10), and they are scalable to reach most employees (P2,6,7). However, some SAMs (P3,9-11,15) were critical about it, often did not see the advantage, and some would even like to get rid of it: \removedForSpace{\contextquote[P10]{I hate phishing just to say. But I do have to run it because that's part of the job and I can't get away from it.} or}\contextquote[P9]{Phishing, I try to minimize it where I can, but must, so a regulatory default actually with us, must be made}. 
Others (P4-8,12) believed that phishing simulations would indeed bring a behavior change among employees. \removedForSpace{: \contextquote[P5]{We sent out our first phishing emails in the Group in 2020. We had 100 calls a day to the help desk or IT support. Now [...] we had seven calls.}.} 
\removedForSpace{P7 even explained that employees would love phishing simulations and would demand emails that are harder to spot: \contextquote[P7]{After up to 15 reported emails, they still have the possibility to get into Spicy mode, which are already difficult emails.}.}

Classical e-learning (\contextquote[P1]{very classic web-based training courses for security, data protection, and physical security.}) was common in the SAMs' organizations (P1,2,4,5,7-10,12,15). Only a few SAMs sought personal face-to-face contact with employees (with the most important face-to-face meeting between employees and security being onboarding). P11 explained why this would be important: \contextquote[P11]{What keeps people secure will absolutely not be the cybersecurity awareness training they had to do on onboarding. What it will be is the fact that during that onboarding they met the cybersecurity team or somebody from it, or they had an onboarding before they were even in the company they were maybe taken through}. 
Gamification and serious games, as a way to improve the acceptance of security awareness activities and as a new promising trend, were exclusively brought into the discussion by the DACH-based SAMs (P1-5,7-9), \eg, \contextquote[P4]{we are still looking for it to be somehow interesting and interactive or to include gamification.}. 

The third most important activity (P1,2,4-6,8,9) was described as active communication into the organization through intranet articles, blog posts, emails, and messenger platforms: \contextquote[P6]{The platforms that we serve both internally [are important], whether it's a social media channel, like Yammer or whether it's a SharePoint, where we sort of prepare all, yes, everything around cyber security.}. P8 was critical about the range they can generate through such communication: \contextquote[P8]{often you can't necessarily reach the masses with videos, blog posts, and linking or advertising your own content.}. \removedForSpace{In P2s' organization, they seemed to have found a simple solution for this: \contextquote[P2]{our Intranet page is a compulsory page, everyone subscribes to it, because of this the topics all arrive before and what we do for example is, we try to keep the articles as short as possible}. This then appears to negate the need to `communicate' with employees, by pushing updates directly to them.}

Security Awareness Days, Live Hacking Events, and On-Site Training were mentioned by some SAMs (P1-4,8,12) but always with the exception that such activities would not scale up: \contextquote[P8]{The share of live training is by far not as large as the share that we try to bring to the people via online content, so to speak. Because we only do such live training courses on request.}. Only P4, P5, and P7 talked about whole campaigns that would include material like posters, or flyers: \contextquote[P7]{They don't look at the intranet, but if there's a postcard lying around, they're much more likely to take it than something digital.}. P4 was the only one that reported that they would try social engineering techniques on their colleagues on-site, \eg, by placing flash drives in the offices or checking the screen locks of Desktop computers.

\subsubsection{Topics} 
\label{results:topics}
Employees were trained on a mixture of threats, attack models, defense strategies, and concrete behavior. Topics were driven by a mix of `basic' technologies such as passwords, and emerging concerns such as remote work and ransomware -- focusing more on the explanation of attacks, than specific behavioral responses to threats.
Password policies, social media security, the darknet, social engineering, tailgating, flash drive security, ransomware, security during mobile work, and data classification were some of the examples the SAMs gave, about what they trained the employees on. Regarding those mixtures of topics, P3 observed \contextquote[P3]{that people are actually always trained on strange topics. So there's usually a quick guide or a collection of topics on phishing, passwords, access, or something like that, i.e. clean desk, and then maybe remote work. But these are completely different categories. You have phishing, which is an attack technique. You have passwords, which are a security measure. You have remote work, that's a place. Clean Desk is another security measure.}.

Seven SAMs (P1,2,5,7,9-11) found it important to relate topics to the private lives of the employees since this would create much more attention for security, \eg, \contextquote[P11]{If I did a workshop on phishing, nobody will come. But if I want to do workshops on dating app security, using Grindr safely, you know, how to send nude photos correct, [...] they love it, you know.}.

\subsubsection{What Informs Awareness} 
\label{results:whatInformsAwareness}
The selection of topics, activities, and awareness strategies were influenced by multiple factors. The biggest influences came from within the organization of the SAMs, highlighting the necessity for awareness strategies tailored to specific organizational needs. Where Haney et al. note the involvement of the IT team~\cite{haney2022approaches}, we find that this can result in sporadic content being created, and in a manner disjoint from regular training. New topics are then addressed in small ways by email at first, potentially long before they become part of regular, structured training. Addressing new threats becomes advisories, distinct from the established training, unless the SAM has resources to quickly make training for newer concerns.

Six SAMs (P4,5,7,8,12,15) stated that their organizational security policies would influence their activities and the content of learning materials: \contextquote[P5]{So 95\% of the policies are integrated into this e-learning}. Some explicitly distinguished between their more sophisticated activities (like live hacking or gamified learning), and some mandatory e-learning only required to explain security policies. 

Three SAMs (P1,6,7) explained that they would send out communication on short notice if topics would come up in the news: \contextquote[P6]{if incidents are picked up by the media, then it is often the case that we try to jump on it and, for example, pick up on a new fraud scheme such as multi-factor authentication phishing.}. The SAMs reported that sometimes internal stakeholders influenced the activities, such as management (P1,2), the communications department (P1,9), the security team (P4,6,8,9), or the CISO (P4,10), \eg, \contextquote[P9]{With the Security Operations Center, we have a monthly appointment where we look, are there any points that we want to pass on [to the employees] each time}. Moore et al.~\cite{moore2016identifying} noted a similar phenomenon for CISOs, needing to respond to the concerns of higher-up managers.

P2 and P10 reported adapting their content to the wishes of employees. Only P4 and P8 would create content around technical updates. \removedForSpace{: \contextquote[P8]{if any new functions or something like that is activated [...] there might be a risk or a problem if it is handled incorrectly. Then, of course, we make a video about it or a communication about it as quickly as possible.}.} 
For P4, P7, and P8 the changing threat landscape leads to adaptation in their security awareness communication: \contextquote[P4]{At the moment there is a huge vishing campaign by some attackers, especially in our sector. And it's happening relatively quickly. So the topic came to me, too, but it also popped out at [the CISO].}. Only P7 reported that a previous incident at their organization influenced the awareness activities. \removedForSpace{\contextquote[P7]{an incident [...] which led to the need to take a more in-depth look at the issue of removable storage devices. And so we reacted to that and turned it around and said, 'Hey, we're doing the issue now.'}.}

\subsection{What is (Good) Security Awareness?}
\label{sec:results-whatisgood}
We explicitly asked the SAMs to define security awareness. What we found was that interpretations were as diverse as the background of the SAMs, covering all aspects of \enquote{Knowledge, Awareness, Training}, with some SAMs explicitly negating some of those aspects. Others were hoping for a redefinition of the term and the whole field. For P2 it was important to distinguish awareness from training: \contextquote[P2]{But actually we are awareness and the awareness [team] does not execute training [in theory]}. 

For others (P1,3-6), training was equivalent to awareness or at least the core of it: \contextquote[P3]{When someone says 'I'm doing security awareness', [...] you train and you educate and you adapt processes.}. P7 set awareness equal to phishing training and P10 reported having heard such from others: \contextquote[P10]{They thought phishing training was awareness. And I had to burst that bubble and basically say no like phishing is such a [...] small element of awareness.}. P3 and P10 acknowledged that it is hard to actually state what awareness really means: \contextquote[P3]{when someone asks, 'Hey, do you do security awareness?' and the person says, 'Yes.' Then that can mean anything from, I wrote an email yesterday and next year I'm going to do this again, [...], to, I'm going to do it like company X [implementing a full security communication strategy]}. 

Multiple SAMs stated that they disliked the term security awareness for what they did, and would like to use alternatives, like security communication (P3), security training (P5), security behavior change (P10), security human factors (P11), security user experience (P13), security culture change (P14), and security engagement (P15). This is notable given that `security awareness' as a term when unpacked typically represents awareness, education, and training. Those SAMs see a shift in the field of security awareness.

\subsubsection{Goals of Awareness} 
\label{results:goal}
One might assume that increased security levels would be the goal of any awareness activities. We found that this assumption might not be true. Awareness as practice seems for the majority of our participants to be about reaching employees and eliciting visible responses.
Only a few SAMs (P3,5,9,10,12) stated that the actual behavior change of employees was indeed a goal of their work (\eg, \contextquote[P5]{the premier class: that the employees actually act securely in our interests.}) or more generally that they wanted employees to become aware of security. 

More important to the SAMs (P1-3,5-8,10) was that awareness itself reached the employees (\eg, \contextquote[P1]{if more than 10 percent of people have actually read the article or reacted in some way, then it's a success}) and that employees would start talking about security and the training they received.
Most participants (P2,3,6,8-10,12,14) saw the creation of visibility for security at the core of awareness. \removedForSpace{\contextquote[P10]{it is really about the purest form of awareness: building recognition and visibility.}.} In the interviews, the SAMs regularly came back to numbers that showed how they achieved greater reach with their activities, like the number of participants in training and live hacking events, or the number of followers to the intranet: it was a prime topic they wanted to bring up. Haney et al. also remarked that US governmental awareness managers saw communication as important \cite{haney2022investigation}, but a complementary NIST report noted a challenge in reaching employees despite using a range of awareness formats similar to those we found here \cite{haney2022approaches}.

When SAMs (P2-4,6,7,9,15) reflected on the awareness goals of their organizations, the most common statement was that the security and risk strategies addressed some human factor, or some human risk and hence required awareness: \contextquote[P2]{So we have defined three main pillars in cyber security, including phishing, and that is the main goal in awareness, to have the human element as well under control as possible.}. 
Only P9 directly connected their work with the actual protection of organizational assets: \contextquote[P9]{The goal is, of course, to be able to better protect our business data as well as our customer data. In other words, simply to meet the level of protection.}. 
Some comments indicated that the goal of security awareness might be self-serving, aiming to find users to train, and not about evidencing secure behavior: \contextquote[P4]{Sometimes there is a campaign that just doesn't succeed. So we did spear phishing, we were too secure. So the emails didn't get in, my people were already too aware. [...] we paid so much money [for the campaign].}.

\subsubsection{Measuring Success} 
\label{results:measuring}
SAMs measure training response but are not sure if they are using the right measurements. They are then not directly measuring security behavior, but instead engagement, where if employees are engaged they are willing to invest themselves in security. Haney et al. noted dissatisfaction with measuring training completion and reliance on correlations with incident reports~\cite{haney2022approaches}.
Awareness activities need confirmation -- from the SAMs and other involved stakeholders. This is why all SAMs reported to have some form of measurement in place. Such measurements could be carried out before, during, or after an activity. The measurements were roughly aligned with the stated goals of awareness (see Section~\ref{results:goal}). However, the SAMs seemed to be uncertain whether the currently available measurements would indeed help with this.

The primary measures of success for the SAMs were (I) engagement- and click-rates on their content (P1-4,6,8,9, \eg, \contextquote[P4]{What are the access rates on the volunteer awareness blog? Yes, I measure that, how successful we are.}); (II) the positive feedback they received (P1,2,5-9,12, \eg, \contextquote[P2]{What's also important with us is that after the sessions, that we then ask, `What did you think of the talk?'}), and; (III) the reporting numbers from phishing simulations (P2,5-7,9,10,12,14,15) \removedForSpace{, \eg, \contextquote[P2]{those classic reporting rates are going up, click rates are also going down.}}. P3, P6, and P14 also used classical metrics from marketing: \contextquote[P6]{An important indicator for us is the NPS, the Natural Promotor Score, where we have a number that indicates how satisfied our colleagues are with our formats.}. Questionnaires to collect feedback were used by seven SAMs (P4-6,9,13-15). 

Some participants (P1,4,8-10,11,15) were transparent that they did not know how to measure the success of their work in a meaningful way, or that their current measurements were limited: \contextquote[P1]{We have already had many discussions about how to measure security awareness, whether we can somehow measure that people are aware of it. But somehow we haven't found any good solutions yet.}. Especially when it comes to the reporting rates from phishing simulations some SAMs (P9-11,15) were skeptical: \contextquote[P9]{How click-through rates change, I think that's what most people mention first. And where they also say, I'm measuring behavior like that, where I'm like, 'No, you're not.'}. 
Another problem identified by P3 and P10 was that measured results would not inform any decision about behavior, but instead about the success of the communication drive itself. \removedForSpace{P10 even experienced that a vendor removed analysis from their dashboard, since customers would not use them: \contextquote[P10]{The vendor [...] decided because nobody used the metrics, and used the dashboards, they actually dumbed down the dashboarding [...] They basically said to me: 'Well nobody else uses it. So why are we over-engineering the idea of the analysis of data when people just aren't using it?'}.}

\subsection{External Influences on Awareness}
The SAMs' work was influenced by factors outside of their direct control, which we explore in the following.

\subsubsection{Vendors}
\label{results:vendors}
We highlight here that SAMs have something of a dependence on vendors, beyond interaction and collaboration as noted elsewhere~\cite{haney2022approaches}. Reflecting the trend of a growing market for security awareness, 13 of the SAMs reported working together with security awareness vendors. They were brought in to help with designing the material, providing e-learning platforms, and managing phishing simulations (\eg, \contextquote[P5]{with the phishing simulation, I also write the email myself and the format and everything [...] and the service provider takes it and builds it into their simulation and that's it. And then sends it out.}), or even developing whole communication strategies. Vendors were either phishing simulation and security training vendors, marketing agencies, or consultants. Hence the influence of the vendors varied heavily from taking over parts of the SAMs' job of planning awareness strategies to being simply suppliers for content.

The primary reason to bring in a vendor was cost reduction, or better: a lack of human resources to do the job without the help of vendors, and hence a reduction of workload for the SAMs, \eg, \contextquote[P2]{We used to do that [content creation] ourselves, fortunately, we bought it from a vendor.}. This seems unsurprising since our participants were responsible (solely, and in a few cases in small teams) for the awareness of thousands of employees. Doing their jobs without vendors might therefore be impossible.

However, a few SAMs questioned how helpful the vendors were in the end. P7 for example complained that international vendors were unable to adapt their content to the needs of a small country like Switzerland. \removedForSpace{: \contextquote[P7]{e-learnings, the issue that is there, really is that does not customize to the extent that it fits Switzerland. I have providers from Germany, and providers from America, are fine. Providers from America, then the German does not fit. There the translations are just wrong.}.} For P5 it was quite important that the concrete training content was created by themselves: \contextquote[P5]{I don't have a portfolio of service providers where I have [many] different measures that I put on a shelf, where I go into the shelf and pull them out, nope. We develop the measures}. For P4 the opposite was the case: \contextquote[P4]{My service provider usually still does some of the writing when I have a blog article}.
Four SAMS (P3,10,11,13) stated that vendors are not worth their money: \contextquote[P11]{No it's not worth it. Make it yourself. There's plenty of educational apps, that you can use which are free or very cheap. And you can make something a lot more engaging and you're going to save yourself 75,000 pounds, you'll raise or whatever.}. \removedForSpace{, or \contextquote[P13]{You can legitimately ask, why are you wasting money?}.} \removedForSpace{They explained that security awareness vendors were solely as successful as they are because they are good at selling shiny and good looking products and so far no one would have asked for the evidence that their product would work, beyond checking boxes, \eg, \contextquote[P13]{And unfortunately, we live in a world where you know, if you can sell something shiny, and it looks good, and it can convince you that it's going to tick your box, and executive looking at our world [will buy it].}. \removedForSpace{P11 added that vendors are also loud and present in the security communities and in social media and hence opinion leaders in the field of security awareness: \contextquote[P11]{people get their understanding from what vendors say. And it's almost like the cool kids in the classroom, you know, a viral on social media. If there's enough people saying it, it's like the emperors' new clothes.}.}}

\subsubsection{Regulations \& Audits}
\label{results:regulations}
We found that Regulations interfere with how SAMs are inclined to manage security awareness.
Since multiple security regulations implicitly (\eg, ISO27001) or explicitly (\eg, BSI Basic Protection) require awareness and training for staff, we were interested in the SAMs' attitude towards regulations. We found differences in the way that some SAMs would come up with the topic themselves early in the interviews, while others would only talk about it after we brought it up. However, with few exceptions, regulations (and the subsequent audits) interfered with most SAMs (P1-6,9,10,12,15) activities. The SAMs talked about this rather abstractly and gave few concrete examples, \eg, \contextquote[P9]{It's actually a discussion I had with an awareness colleague in the community once who said, 'You don't have to phish.' And then when I looked at what FINMA [banking regulation] had written in our review, it explicitly says, 'You have to phish.'}, or \contextquote[P10]{one of the biggest changes of the requirement for the new PCI [industry regulation] it specifically says that you have to teach them about social engineering and phishing.}.
\removedForSpace{P6 explained how they would be ahead of a new upcoming regulation with their activities: \contextquote[P6]{But the fact is that NIS-2 now explicitly requires training and awareness for the first time. This means that we are actually anticipating a regulatory requirement and are making awareness a mandatory task.}.} 

The SAMs denied that regulations would be a primary driver of why they and their organizations do awareness, \eg, \contextquote[P4]{Audits is a necessary evil for everything, but I don't think that was the core point of why we implemented that, but we really want to be secure.}. Some SAMs were proud that the auditors rated their awareness as positive (\eg, \contextquote[P2]{So the audit also always says, 'Our awareness is great.'}). However, no SAM described the regulations as helpful for their own work, \eg, for generating the necessary attention. Even more, the SAMs were critical of influence of regulations on security awareness. They would often demand a type of awareness that would not help their goals: \removedForSpace{\contextquote[P9]{Phishing, I try to minimize it where I can, but must, so a regulatory default actually with us, must be made.} or} \contextquote[P10]{one of the biggest changes of the requirement for the new [regulation] is that you have to teach them. [...] why am I teaching somebody in a retail store, who has no access to the computer and only a pay machine, why am I teaching them about phishing? They don't even have an email address.}.

\subsection{Interactions and Frictions With Others}
\label{results:power}
\label{results:alliances}

Interactions with several functions in the business act to limit or dictate SAM activities, namely regulators, and communications teams, which some SAMs felt negatively (\eg, \contextquote[P1]{the communication department is very obstructive anyway, I have to say quite honestly}), and their reports/managers, as well as technical teams. 

Most SAMs were convinced that they had a lot of freedom in their decisions, when it comes \eg, to the content of training or the choice of an appropriate vendor.
However, they noted that when their experiences and accounts of security in practice clashed with the expectations of the technical IT team, they would be overruled.
Sources elsewhere note that most SAMs are part-time and otherwise in technical security roles~\cite{SANS.2021.AwarenessManagerReport,SANS.2022.AwarenessManagerReport}, whereas our participants are an exception. 

\subsubsection{Interactions With Technical Teams} 
\label{results:cybersecurityistechnical}
The technical side of security played an important role in the SAMs' daily work, with most SAMs (P1-5,9,11,12,15) regularly working together with the technical security teams, \eg, security operation center teams, security infrastructure, or internal penetration testers. However, limitations were described in the relationship with technical teams by multiple participants (P3,8,10-12,14,15) when there were disagreements, as they reported that security awareness and SAMs are not acknowledged as equals by their counterparts in the more technical security teams, as \contextquote[P10]{they're the ones that will often say well you can just train people or you can just tell people}, and \contextquote[P12]{the value of [the] security awareness and training team can get better no doubt.}.

Some SAMs also explained that if those teams tried to influence their work, they would often not consider updated threat models, \eg for P10's organization when technical teams insist on teaching employees about secure WiFi networks, which is of less relevance when a VPN is always activated on all employee mobile devices, \contextquote[P10]{Just because it's written down doesn't make it true. [...] and of course, I'm just an awareness person. So what do I know about these things?}.
This all points to a misalignment between the expectations of technical security teams and SAMs. SAMs are outnumbered and when there is doubt about topics, the technical side is more established and \contextquote[P10]{believed more.}

There seems to be a disconnect in the mission of different teams. While our SAMs are in constant contact with the technical security teams, they do not interact and align their activities: awareness content is rarely adapted to technical security changes. Even more critical, the technical security teams want to decide on their own what to communicate to employees, and hence override the SAMs' awareness strategies.

\subsection{Interactions with Employees}
\label{sec:res:employees}
As the SAMs' efforts were in one form or another directed to employees, it is key to understand their perception of employees.
Some form of relationship between the SAMs and the employees was key for every participant, but on different levels: for some (P1,7,10,11,14) it was at the core of their job to communicate with employees face-to-face or via email (\eg, \contextquote[P14]{The security awareness team has the fantastic opportunity to be the friendly face of security}), while others would like to reduce the number of questions they get from employees (P5,6).
Almost all SAMs told us that customized training for different groups of employees is key for their work -- that training needs to be adapted to the different job roles. Team leaders, blue-collar workers, software developers, and retail employees would need special attention -- with the distinctive needs for e.g., developers and office-based workers being explored in previous research~\cite{beautement2016productive}.

Four SAMs (P6,12,13,15) worked with Security Champions~\cite{tahaei2021privacy,haney2017skills,haney2017work,haney2018itsdull,Menges.2023.CaringNotScaring,gabriel2011selecting,becker2017finding} or Security Ambassadors. \removedForSpace{:\contextquote[P6]{that is a network of colleagues who are involved in cyber security. Who has an affinity for the topic and who are also considered contacts for the business. And this is a community that we also help to develop.}.} Here, P13 described security champions as a way to close the (communication) gap between employees, security teams, and security awareness teams: \contextquote[P13]{And we're seeing success in that compliment that with the security champions or advocates that were people, you know, people on the ground, are able to cascade that information and give you feedback to be able to mature your program.}.
Looking back at the implemented awareness activities (see Section~\ref{results:activities}) we can see that for most SAMs a closer communication and tailored training would be desirable, but these do not scale up for thousands of employees.

\subsubsection{Expected Effort, Security Friction, \& Usable Security} 
\label{results:securityfriction}
The SAMs were well aware that security is only a secondary topic for most employees, and they often stated that security training should not take too much time from them (in contrast to many CISOs~\cite{Hielscher.2023a,Hielscher.2023b}). There was, however, something of a contradiction, in that our participant SAMs sought ways to increase engagement and provide more training, and were looking to make training mandatory, simply to increase otherwise low engagement rates, \eg, \contextquote[P7]{Onboarding in particular is to be mandatory. The phishing simulations will soon be mandatory.}. Hence, it was not that the secondary costs of security awareness were ignored, but were discounted when it came to making time for training.
First and foremost the UK-based SAMs described their (desired) role as being aides for the employees. They looked beyond training and were more concerned about ineffective security processes that caused friction for employees and how they could help to tackle this. P15 explained how their task evolved over time and how they realized that security communication can not fix broken security processes: \contextquote[P15]{We now try and understand why it's happening and what's happened [...] what comes out of it is things like there's a process that is broken and there's no other way to do it [...] and comms would never fix. So we work quite closely with security operations on that to try and find those human risks rather than, you know, just putting up comms.}.
This was an outstanding example, and other UK-based SAMs described this as a task that they and other SAMs should look at, rather than as something that currently is within their job descriptions. \removedForSpace{P13 explained how SAMs would in his opinion misinterpret their role: \contextquote[P13]{almost all of them [SAMs] don't realize that minimizing the burden and making it so easy, it should be part of their role. A lot of them see it as a calming engagement, activity, survey, [...] That's not real work. That's busy work. That doesn't change anything. It's just a charm.}.}

The DACH SAMs showed less thinking in this way. P2 even denied that it would be their task to consider the burden security creates for employees: \contextquote[P2]{as an awareness manager, I don't really care. So what you want to do with your data is a nice idea at first, but we have basic principles and rules. So we say, ``Okay, these are the rules that exist in the company and that you have to follow.''}. While all SAMs were aware that some security policies are hard to follow, they had few ideas on how to tackle these and did not describe it as within their roles. 
Some described the training as a solution for missing usability, \eg, \contextquote[P1]{At the same time, we try to train that e-mails should get encrypted. That works less well because, of course, that's something you have to do actively. But that's something we try to teach.}. Here the SAMs showed the same thinking as CISOs, who see security training as a quick fix to unusable security procedures~\cite{Hielscher.2023a,Hielscher.2023b}.
P5 explained that their organization had a usability team that the head of IT and the CISO were part of, but their work was not related to his own. Similar disconnects between security, software, and usability teams have been documented elsewhere~\cite{caputo2016barriers}.

\section{Discussion}
\label{sec:discussion}
Here, we discuss our findings, with regards to our \hyperlink{researchQuestions}{research questions}, followed by recommendations for research and industry.

\subsection{RQ1 -- The Definition of Awareness}
Regarding RQ1, we found that awareness was regarded as a combination of tangible activities and material delivery (\ref{results:activities}), and ongoing engagement for visibility (\ref{results:goal}).
The term \enquote{security awareness} can mean different things to different practitioners, be it about training engagement or pure dialogue with employees. This contrasts with an under-specification of what \enquote{awareness, education, and training} actually is in practice. This is not a new concern and has already been discussed at least as far back as 2005~\cite{furnell2005organizational}. 
We found that participants filled their role with work that they justified relative to security, but seemingly according to their own rationale. The role of the SAM also differed, sometimes including the creation of content, active communication, or more of a management role, navigating different stakeholders. This was informed by regulations (which were seen either as a guide on content, or a hindrance in terms of being seen as overly prescriptive).

What we have found is that where existing research notes that training may be onboarding or regular (often annual) training, there is another activity of regular advisories on emergent threats. The tension here, noted by participants, is that technical teams may dictate these (fragmented) advisories and that senior managers and regulators may dictate very specific expectations which are not checked for whether they map to the organization's context.

There is then a \enquote{void of specification} as to what security awareness should be on a day-to-day basis, that our participants filled in a self-directed manner. SAMs are filling the void with seemingly good things, and what is considered state-of-the-art by their peers. There is then fragile success, that rests on technologies being usable and time being afforded for employees to enact security tasks, so that employees not knowing or doing the behaviors can be the only possible reason for non-secure behavior. This then presumes that awareness initiatives are the only solution needed to address the persistence of non-secure behaviors.

\subsection{RQ2 -- Managers and Employees}

Based on our findings, it seems that employees do not complain about awareness activities (\ref{sec:res:employees}), so SAMs (and CISOs~\cite{Hielscher.2023a}) can assume awareness must be working. Other scholars have already shown that support moves to self-help, resulting in invisible workarounds~\cite{Kirlappos.2014}, and awareness is knowingly configured so that the influencer can claim success: most employees pass phishing tests over time (\ref{results:measuring}). 

What emerges is a kind of `inverse usability' check, where it is not that SAMs check that security solutions are first usable, but instead make themselves available to hear about difficulties with security solutions. This is similar to providing an abundance of training materials, and being available if employees struggle. SAMs see engagement as important, so that employees can signal if they need information or extra help, but reliance on vendors and restrictions on resources mean that SAMs can only repeat or intensify the same assistance. The common approach is to ensure that employees know who to contact if there are problems, rather than investing upfront to ensure there are easier security solutions.

Ultimately the SAMs can define goals themselves and how to measure their success. NIST standard 800-50~\cite{nist800-50-2003} describes that awareness needs to fill a gap between existing security knowledge and identified needs. Reflecting on this gap, our results suggest that while SAMs consider specific content for different employee groups, there is little that moderates whether employees' specific needs are being met. Instead, a range of materials is communicated to employees coming from all groups and professions, rather than crafting content to specific needs. This is not necessarily the SAM alone -- topics may be defined by regulations (\ref{results:regulations}), internal technical teams, and `market standard' approaches/products used. SAMs then communicate it and determine how employees engage with the content. This represents an approach of providing a maximum of content and seeing how it lands with the audience.

\paragraph{Missing Ownership}
Despite a thriving field of usable security research, in the organizations described here, no one feels responsible for implementing usable security, bringing awareness and technology together, and taking the burden security creates for employees into account: not the CISOs~\cite{Hielscher.2023a}, nor security consultants~\cite{Hielscher.2023b}, nor SAMs (\ref{results:securityfriction}). No one has ownership of unusable security tools and policies. The remit -- let alone the power -- does not exist to put awareness and technology design together as advocated by prior research~\cite{beautement2008compliance}. 

If the job of SAMs is to only get the message out, it relies heavily on technology already being usable, since this is not being tested and is not something in the power of the SAM to change (if they were to see it causing problems for users). In our study, guaranteeing secure employee behavior is costly to measure, but also does not seem to be within anyone's remit. This raises questions as to how `usable' security solutions and associated training reach organizations in the first place. Studies aiming to quantify the costs of unusable security are rare~\cite{herley2009so,beautement2016productive}, but are needed to make a case for a usable security mandate within organizations.

Early research in Human-Centered Security~\cite{flechais2005divide} indicates that the research community was focused on building more usable tools; more recently, some 20 or so years later, in organizations, the concerns of (I) having secure tools and (II) effective use of secure tools now run in parallel rather than in tandem, and do not interact.

\subsection{RQ3 -- Success of Awareness}
The SAMs are not relying on a direct measure of the practice of secure behaviors (\ref{results:measuring}). Instead, awareness engagement itself, rather than the security of behaviors, is the typical measure; this has been seen in the past but in terms of \eg, how to automate away the effort of total coverage~\cite{alshaikh2020developing} (how to produce content efficiently to reach maximum engagement). We found that some SAMs were engaging in varied and nuanced ways, not just with training materials but also with a general openness to interactions that boosted engagement. A positive attitude to security was seen as an important measure, through any channel. 
SAMs then use the feedback of employees and reporting numbers in phishing simulations as proxy measures, to assess how the employees perceived their activities. Those activities are mainly built around some form of training \& communication. The SAMs' job is about bringing messages out to the employees as much as it is about the message itself. 

In this sense, the SAM's engagement activities prepare the ground, to be ready to capture feedback if something in the mix of materials is not landing well with employees. Our SAM participants were watching for signals that: (I) the material was acknowledged, but also; (II) whether there were problems indicating that needs were not being met. This was epitomized by P7, who thought onboarding was just as important for making the cybersecurity team known in case there were issues down the road for new employees.

\paragraph{Awareness as a Proxy for Secure Behavior}
While security awareness is about getting employees to follow security rules -- at least following norms like ISO27004~\cite{iso27004.2016} or BSI Basic Protection~\cite{bsi.2022} --, no one is actually measuring whether they are following the security rules. It is simply assumed that awareness will reinforce secure behavior; awareness then becomes a proxy for secure behavior. This works most aptly for anti-phishing, but anti-phishing measures are co-opted for a range of reasons including to measure a behavior~\cite{Hielscher.2023a} (but technically the simulated phishing emails, rather than actual malicious phishing emails).

What is dangerous about such an approach is that it risks assuming that if awareness or engagement is increased, secure behavior would increase.
These efforts could be successful in increasing awareness alone, but do not resolve usability issues or inadequate policies. Awareness practice is driven as if technologies and policies are unquestioningly usable. This then has parallels to how the job of making software usable is treated as something `part-time'~\cite{caputo2016barriers}.

We saw also in~\ref{results:goal} that practitioners felt there would always be something that employees would need to learn; it is not completely clear how to separate this from awareness delivery being a person's job -- being seen to deploy awareness materials as a measure that perpetuates a belief that employees will never know enough. It is then unclear if awareness as a discernible role means that SAMs are incentivized to treat employees like they never know enough; can a SAM expect to be paid if they declare that all staff know all the training?

This is all to say, that there is the unchecked provision of support materials into the organization, which satisfies the need to be seen to be meeting regulatory needs -- what is checked is how well the materials are digested by the workforce. This also has the indirect benefits of: (I) the organization (including SAMs) (externally) looking like they are doing their job, and; (II) avoiding the cost of measuring the aforementioned gap defined by NIST 800-50. It is arguably cheaper to meet a complete set of needs than it is to identify a specific set of needs, hence over-communication, with parallels to advice over-production~\cite{Neil.2023,redmiles.2020,reeder2017152}.

\subsection{Lessons For Researchers}

Here we present recommendations based on our outcomes, framing opportunities for further research.

\paragraph{The Not-So-Easy Task of Communication}
More research needs to focus on communication and this less tangible part of a SAM's job (\ref{results:goal}). Some research in security management has considered communication, through social marketing~\cite{ashenden2013can}, dialogues between IT teams and employees~\cite{ashenden2016security}, and advice construction~\cite{neil2023comes}. The research community should make space for understanding how communication efforts link to security artifacts in organizations, such as policies and controls.
More concretely, the metrics proposed by Ashenden and Lawrence in their `Security Dialogues' work \cite{ashenden2016security} could be adapted (as has been proposed for SME conversations with IT providers  \cite{parkin2021change}), for instance not only recording how often help or a conversation is sought (as our SAMs already reported), but finding a formal mechanism for documenting this evidence. Going beyond the `Security Dialogues' work, this could also include logging when user feedback influences a change in policy and training (as noted in other user engagements, e.g., \cite{beautement2016productive}). Otherwise, the freedom of SAMs appears to rely on their drive and ability to convince others.

\paragraph{Capture the Splintered Nature of Security Awareness} 
We argue that researchers using the term `security awareness' -- or indeed, any one term -- in their own work need to specify what they mean by security awareness (\ref{sec:results-whatisgood}). Aside from the basic distinction between security education, training, and awareness, we found that SAMs deploy strategic campaigns, send emails to respond to new issues, reach out to employees, etc. We suggest distinguishing between (I) short-term \textit{training or awareness-raising interventions} like phishing simulations, (II) long-term \textit{strategic educational efforts} where different training modules are built on each other, over months or even years, and (III) \textit{security communication}, where \eg, employees are simply informed about a new attack via a newsletter or email. This distinction reflects the categories of activities we identified among our participants. Without such distinction, it is more difficult to compare findings or categorize them across studies. This would require consideration of \eg, reactive alerts as part of the security sensitization and awareness activities of organizations, rather than focusing on assuming or focusing only on either a fixed regular training package, or an assumption that all security awareness in organizations is qualitatively the same.

\paragraph{Explore Alternative Success Indicators for Secure Behavior} Our participants were generally split, between providing content, and focusing on communication. The former sets up a situation that disincentivizes ever stating that employees have enough information -- otherwise the SAM risks not appearing to be doing anything. The latter was seen as having value but relies on the standing of the SAM to argue the case to non-SAMs (who are informed by the limited specification of what a SAM is expected to do).

The success of awareness is measured through indirect indicators, a common challenge in organization cyber-risk management~\cite{woods2021systematization}. Secure behavior measurement would require instrumentation of the entire digital estate, or at best measurement of very narrow behavior definitions (which may be rigid and would need adaptation to particular contexts, as with e.g., SeBIS~\cite{egelman2016behavior}). For some behaviors it would seem straightforward to measure a behavior change, \eg, the usage rates of password managers, MFA, or passkey, or the number of screen locks, the usage rates of cloud encryption tools, regularity of software updates, etc. However, these would require instrumentation of end-user machines (which is an issue if, as we have found, technical teams can `over-ride' SAMs to dictate the importance of technical issues).

Regarding phishing simulations, it would be important to measure (I) whether employees reject legitimate e-mails out of fear (as a suggested negative side-effect~\cite{Volkamer.2020}), and for larger organizations (II) whether the number of successful real-world phishing attempts is reduced after a phishing simulation. There is scope to explore the `inverse usability' indicators noted earlier in this section, especially while the usability of security solutions is determined by vendors. This would require closer interactions with the helpdesk, but also could leverage suggestions from work on `shadow security' \cite{Kirlappos.2014}, to have a team or middle managers log queries from staff.

\subsection{Lessons for Industry and Policy}
Here we summarize what organizations and policymakers could act on to explore improvements.

\paragraph{SAM as Advocate for Employees and Usable Security}
With security teams still having dysfunctional relationships with the rest of the organization~\cite{dasilva2022cyber,ashenden2016security,menges2021security,enisa2019cybersecurity}, SAMs -- as self-described security communicators -- can jump in and fill the gap.
SAMs should be the employees' advocates: the friendly faces of the security team (P14). We advocate a general move to understand how more usable technologies enter an enterprise environment. SAMs could be involved in IT and security technology procurement processes, as advocates representing employees' needs -- then, the representative of the `user' (employee) has sight of products potentially entering the organization, more so than the `user' (security manager) as a customer purchasing the products. A SAM can surface frictions that may appear once a product is deployed, based on their interactions with employees (as has already been hinted at with phishing product procurement~\cite{Brunken.2023}).

\paragraph{Surface Usability Within Assurance Measures}
If there is an underspecification of any measurable approach or goal for awareness, vendors can sell products to organizations without having to bear the cost of proving that they are effective; this shifts the risk and costs to the customer to make it work, as a `negative externality'~\cite{anderson2001information}.
The revision of NIST 800-50 was open for comment around the time of writing
\footnote{\url{https://csrc.nist.gov/News/2023/nist-releases-draft-sp-800-50-rev-1}, accessed \today};
similar efforts to gather experiences from the SAM community as a specification for awareness could help to avoid a market of `silver bullets'~\cite{grigg2008market} -- where neither client nor their users know what works best -- around security awareness, towards evidence-based security~\cite{Hielscher.2023a}. Our participants were dissatisfied with the increasing and prescriptive details of regulations requiring concrete content that risked not fitting the individual needs of their organization. This needs to be tempered, as P10 noted in response to our results, that organizations are too complex to directly measure security behaviors and the connection to security awareness. However, our participants did collect feedback through large-scale tools, where assurance expectations may lift the profile of those tool solutions, to ease the development of feedback mechanisms.

\paragraph{Give the SAM a Stronger Mandate}
We found SAMs were left to fight for what they believed needed to be done for employees to work securely. Regulations could be adjusted to add weight to the feedback that SAMs receive, as an assurance measure. This could be a regulatory mandate, similar to Data Protection Officers and privacy needs, to ensure that \eg, when an employee signals that they find a technical measure unusable, this does not get left unaddressed (for lack of either the employee or the SAM being able to push back against the expectations of technical and senior management teams). Linking to the previous recommendation, this then in turn gives organizations a mandate to match solutions to their specific context and reduces `risk dumping'~\cite{anderson2001information} upon employees to make security work. Vendors do not necessarily have this insight into organizations when they provide solutions (instead relying on what is told to them during procurement, typically by a technical manager).

\section{Conclusion}
Here we explored goals and drivers of security awareness in practice, through in-depth interviews with $n=15$ European Security Awareness Managers. In the absence of well-defined, measurable, evidence-based security awareness -- which would lead to secure behavior -- those managers carry out awareness as what they perceive as successful: communication about security that reaches as many employees as possible. The managers were critical of the status quo (\eg, that phishing simulations might not work, but are required by some regulations), but they lacked a clear mandate for change.

We conclude that researchers need to clearly distinguish different aspects of the over-used term of security awareness, need to include the challenge of creating engagement in their work, and need to help identify clear measures of success for awareness activity -- and hence move forward to evidenced-based security awareness. Organizations need a clear mandate for usable security processes over communication alone, with awareness managers being natural holders of such a mandate. 

\section*{Acknowledgments}
We thank all participants for their openness and the time they spent with us. We would like to thank Julian Becker and Tatiana Mikhaylova for their help with the transcription. The work was supported by the PhD School "SecHuman -- Security for Humans in Cyberspace" by the federal state of NRW, Germany, and partly also by the Deutsche Forschungsgemeinschaft (DFG, German Research Foundation) under Germany’s Excellence Strategy - EXC 2092 CASA - 390781972.

\bibliographystyle{plain}
\bibliography{Awareness-Manager.bib}

\appendix
\section{Interview Guide}
\small
\label{app:InterviewGuide}
\renewcommand{\labelenumii}{\theenumii}
\renewcommand{\theenumii}{\theenumi.\arabic{enumii}.}

\paragraph{Intro + Role}

\begin{enumerate}\setlength\itemsep{0.1em}
    \item Please describe your role
    \begin{enumerate}\setlength\itemsep{0.1em}
        \item Probe: Are you part of a team or are there others also responsible for managing Security Awareness?
        \item Probe: What are your duties and responsibilities in your current position?
        \item Probe: Is there anything in your role that is not Security Awareness?
    \end{enumerate}
    \item How does Security Awareness fit with your role?
    \item What to you is Security Awareness? What is not?
    \begin{enumerate}\setlength\itemsep{0.1em}
        \item Probe: are there any specific `building blocks', such as Awareness, Education, Training, Policy?
    \end{enumerate}
    \item Demographics (if not already covered): How many users does your work apply to in your current role?
    \item Demographics (if not already covered): How many years of experience do you have, and how many organizations does this include?
\end{enumerate}

\paragraph{Role Activities}

\begin{enumerate}\setlength\itemsep{0.1em}
    \item What is done in your organization as Security Awareness?
    \begin{enumerate}\setlength\itemsep{0.1em}
        \item Probe: how well is it working?
    \end{enumerate}
    \item How does a typical work week look for you?
    \item What kinds of decisions do you have to make in your role?
    \item What input/information informs those decisions?
    \item Are there any regular barriers to your Security Awareness activities?
    \item Does your work interact with more technical measures? 
    \begin{enumerate}\setlength\itemsep{0.1em}
        \item If yes, how do your job activities interact with the technical measures? Can you give an example?
        \item If no, what is expected of technical measures that make them separate from your job activities?
    \end{enumerate}
\end{enumerate}

\paragraph{Goals of Awareness}

\begin{enumerate}\setlength\itemsep{0.1em}
    \item What is the demand for security in your organization?
    \begin{enumerate}\setlength\itemsep{0.1em}
        \item Probe: what do you see as the justification for your job activities? 
        \item Probe: Are there points where your job activities are assessed by others? (within or outside of the organization)
        \item Probe: is there a predefined state of security awareness that the organization is seeking, or is it defined more by you?
    \end{enumerate}
    \item What indicates success in your Security Awareness role, and how is your success measured/demonstrated?
    \item What is the goal of Security Awareness for you and your organization?

    \begin{enumerate}\setlength\itemsep{0.1em}
        \item Probe: How much of your organization’s security is reliant on what you do?
        \item Probe: What would happen in the organization if the Awareness did not happen?
    \end{enumerate}
    \item How do the Awareness activities in your organization relate to the goals of the wider organization and other functions?
        \begin{enumerate}\setlength\itemsep{0.1em}
            \item Probe: core values, annual business objectives
            \item Probe: what is the relationship between what you do and the organization’s security policies?
        \end{enumerate}
    \item Are there any examples of where you need to react urgently to specific security issues, or is the pace of activities more planned out? 
    \item What would your work be if all employees were working securely and how would that differ from what you do now?
\end{enumerate}

\paragraph{Awareness Tools \& Techniques}

\begin{enumerate}\setlength\itemsep{0.1em}
    \item What techniques, products, or services do you use?
        \begin{enumerate}\setlength\itemsep{0.1em}
            \item Probe: why these? (resources, recommendations, best practices)
            \item Probe: are there any which are more, or less, reliable than others for ensuring a successful outcome?
        \end{enumerate}
    \item What do you produce that goes to employees? To others?
    \item What characterizes a ‘good’ tool or technique for Security Awareness?
        \begin{enumerate}\setlength\itemsep{0.1em}
            \item Can you provide an example of a good solution, and explain why it is good? 
            \item Is it good for you AND good for employees? 
        \end{enumerate}
    \item What resources do you have for Security Awareness? Do your resources change or stay the same over time?
\end{enumerate}

\paragraph{Role of Employees in Security Awareness}

\begin{enumerate}\setlength\itemsep{0.1em}
    \item What do employees do as security behaviors? (list main ones)
        \begin{enumerate}\setlength\itemsep{0.1em}
            \item Probe: how do these behaviors link to your Security Awareness activities?
            \item Probe: are there values, clear rules, or a policy? Or a mix?
        \end{enumerate}
    \item How is the need to behave securely communicated to employees?
        \begin{enumerate}\setlength\itemsep{0.1em}
            \item Probe: explained/justified to employees in your organization?
            \item Probe: by its own importance, by it supporting other key org. goals, etc.
            \item Probe: what is the format?
        \end{enumerate}
    \item Is there anything employees do that indicates that your Security Awareness efforts are effective? If so, what is that and how is it measured?
        \begin{enumerate}\setlength\itemsep{0.1em}
            \item Probe: do you know what employees do? How sure are you, and how sure do you want to be?
        \end{enumerate}
    \item Do you interact with employees in your organization? What form and regularity do those interactions have? 
        \begin{enumerate}\setlength\itemsep{0.1em}
            \item Can you recall any examples of good feedback?
            \item Can you recall any examples of bad feedback?
        \end{enumerate}
    \item Would you say there are any differences in the Awareness/security needs of different employees in your organization?
        \begin{enumerate}\setlength\itemsep{0.1em}
            \item Probe: Any problems that particular users have?
        \end{enumerate}
    \item Are there any aspects of your interactions with employees that are easier than others? If so, why?
    \item If there was anything that would improve the way employees interact with your Security Awareness efforts, what would it be?
    \item Is there anything outside of Security Awareness that would make Security Awareness easier to achieve or more reliable?
    \item Is usable security considered in the security activities the employees have to carry out?
    \item Is there any negative of the Security Awareness on employees you could think of?
    \item As a hypothetical question: if you answered to employees, what do you think they would ask from you?
\end{enumerate}

\paragraph{Outro}

\begin{enumerate}\setlength\itemsep{0.1em}
    \item If you have any other thoughts or ideas, you can tell us now or write/ call us later. You know, a lot of thoughts come quite sometime after an interview happens.
    \item Could you recommend any other Security Awareness Managers we may talk to?
\end{enumerate}

\end{document}